\documentclass[aps,prd,onecolumn,floatfix,superscriptaddress,nofootinbib]{revtex4-1}

\newcommand{\sigsip}{\ensuremath{\sigma_{\chi p}}}
\newcommand{\kev}{\ensuremath{\,\mathrm{keV}}}
\newcommand{\mev}{\ensuremath{\,\mathrm{MeV}}}
\newcommand{\gev}{\ensuremath{\,\mathrm{GeV}}}
\newcommand{\tev}{\ensuremath{\,\mathrm{TeV}}}

\usepackage{amsmath}
\usepackage{graphicx,bm}
\usepackage[colorlinks,citecolor=blue]{hyperref}
\usepackage[caption=false]{subfig}

\begin{document}
\title{Probing Cosmic-Ray Accelerated Light Dark Matter with IceCube}

\author{Gang Guo}
\email{gangg23@gmail.com}
\affiliation{Institute of Physics, Academia Sinica, Taipei, 11529, Taiwan}

\author{Yue-Lin Sming Tsai}
\email{smingtsai@gate.sinica.edu.tw}
\affiliation{Institute of Physics, Academia Sinica, Taipei, 11529, Taiwan}
\affiliation{Key Laboratory of Dark Matter and Space Astronomy, Purple Mountain Observatory, Chinese Academy of Sciences, Nanjing 210008, China}

\author{Meng-Ru~Wu}
\email{mwu@gate.sinica.edu.tw}
\affiliation{Institute of Physics, Academia Sinica, Taipei, 11529, Taiwan}
\affiliation{Institute of Astronomy and Astrophysics, Academia Sinica, Taipei, 10617, Taiwan}
\affiliation{National Center for Theoretical Sciences, Physics Division, Hsinchu, 30013, Taiwan}

\begin{abstract}
The direct detection of particle dark matter through its scattering with nucleons is of fundamental importance to understand the nature of DM. 
In this work, we propose that the high-energy neutrino detectors like IceCube can be used to uniquely probe the DM--nucleon cross-section for high-energy DM of $\sim$~PeV, up-scattered by the high-energy cosmic rays.
We derive for the first time strong constraints on the DM--nucleon cross-section down to $\sim 10^{-32}$~cm$^2$ at this energy scale for sub-GeV DM candidates.
Such independent probe at energy scale far exceeding other existing direct detection experiments can therefore provide useful insights complementary to other searches.
\end{abstract}

\date{\today}

\maketitle

\section{Introduction}\label{sec-intro}
The existence of dark matter (DM) revealed by cosmological and astrophysical observations through its gravitational effect has been firmly established.
However, the particle physics nature of DM remains elusive despite tremendous experimental efforts over the past decades with both the direct and indirect detection methods.
In particular, the ton-scale underground detectors probing the interaction between DM and atomic nucleus set very stringent limit on the DM--nucleon cross-section down to $\sigsip\sim 10^{-46}$~cm$^2$ for DM mass between 10~GeV$\lesssim m_\chi\lesssim 10$~TeV, approaching the limit set by the neutrino background \cite{Akerib:2016vxi,Aprile:2017iyp,Cui:2017nnn,Aprile:2018dbl}.
Recently, increasing amount of work also started to investigate other motivated DM candidates whose masses are lighter than GeV 
with very different experimental techniques and/or cosmological/astrophysical observation, 
see e.g., 
sub-GeV DM~\cite{Knapen:2016cue,An:2017ojc,Ibe:2017yqa,Dolan:2017xbu,Cappiello:2018hsu,Bringmann:2018cvk,Ema:2018bih,Cappiello:2019qsw,Dent:2019krz,Berlin:2018sjs,Akesson:2018vlm,Matsumoto:2018acr,Berlin:2019uco,Depta:2019lbe,Hertel:2019thc,Alvey:2019zaa,Dror:2019onn,Dror:2019dib}, 
QCD axion DM~\cite{Kahn:2016aff,TheMADMAXWorkingGroup:2016hpc,Co:2017mop,Hook:2018iia,Obata:2018vvr}, 
sterile neutrino DM~\cite{Speckhard:2015eva,Adhikari:2016bei,Arguelles:2017atb,Brdar:2017wgy,Caputo:2019djj,Hofmann:2019ihc,Arza:2019nta}, 
and ultralight Bosonic DM~\cite{Hui:2016ltb,Dev:2016hxv,Fukuda:2018omk,Chen:2019fsq,Guo:2019qgs,Carney:2019cio,Davoudiasl:2019nlo}.

Among those, Refs.~\cite{Bringmann:2018cvk,Ema:2018bih,Cappiello:2019qsw,Dent:2019krz}
pointed out that the collision of the diffuse cosmic rays (CRs) with energy up to a few GeV in our Milky Way (MW), including nuclei and electrons/positrons, with DM in the halo can up-scatter a small amount of DM to velocities far exceeding values carried by the locally virialized DM.
This up-scattered DM component with higher velocities can produce larger recoils for targets in existing DM or neutrino experiments than the local DM and 
gives rise to new exclusions to the DM--nucleon(electron) cross-section for keV~$\lesssim m_\chi\lesssim$~GeV, inaccessible by DM experiments without considering this component. Applications of such scenario to specific models have also been explored in Ref.~\cite{Wang:2019jtk}.

In this paper, we consider the acceleration of light DM by high-energy (HE) CRs 
with energy of $\sim$~few PeV\footnote{We note that this differs from the boosted DM scenario 
in which HE dark particles can be produced 
from decays of heavy particles~\cite{Bhattacharya:2014yha,Kopp:2015bfa, Bhattacharya:2016tma}, 
or annihilations of high energy cosmic neutrinos with comic background neutrino \cite{Yin:2018yjn}.}. The HECRs in this energy range can accelerate DM to very high energy of $E_\chi\lesssim$~PeV.
These ultra-relativistic DM can then masquerade as HE neutrinos when they scatter inelastically with nucleons\footnote{We assume the same cross-section for DM--proton and DM--neutron scattering.} in HE neutrino detectors, such as the IceCube Neutrino Observatory \cite{Aartsen:2016nxy} or KM3NeT \cite{Adrian-Martinez:2016fdl}.
In particular, the deep-inelastic scattering (DIS) of DM in ice or water will predominantly produce cascade events, but not the tracks associated with muons.
As astrophysical neutrinos produced from charged pion decay are expected to have nearly equal flux in each flavor arraiving at the Earth after traversing cosmological distances: $F_{(\nu_e+\bar\nu_e)}:F_{(\nu_\mu+\bar\nu_\mu)}:F_{(\nu_\tau+\bar\nu_\tau)}\simeq 1:1:1$~\cite{Learned:1994wg}, 
one naturally expects that the extra contribution from HE DM masquerading as HE neutrinos can produce an excess in cascade events at some specific energy ranges and angular directions while deviate the flavor ratio from equi-partitioned.   
Consequently, unique constraints on DM--nucleon cross-section at energy range of $E_\chi \sim$~TeV--PeV in the rest frame of nucleon can be obtained, which will be useful when combined with constraints derived at other energy scales \cite{Bringmann:2018cvk,Ema:2018bih,Cappiello:2019qsw,Dent:2019krz}.

Below in Sec.~\ref{sec-dmflux}, we first compute the up-scattered DM flux in this energy range by the HECRs inside our galaxy as well as the DM flux accelerated by the extragalactic HECRs as a diffuse source. 
In Sec.~\ref{sec-result}, 
we compute the expected DM event numbers in different angular bins at the IceCube, and perform statistical analysis using the released HE neutrino data from the IceCube to derive new constraints on DM--nucleon cross-section at energy range of TeV--PeV.
We discuss the implication and conclude the paper in Sec.~\ref{sec-summary}.

\section{Ultra-relativistic DM upscattered by HECRs}\label{sec-dmflux}

\subsection{Galactic HE DM flux}\label{sec:GDM}

Following Refs.~\cite{Bringmann:2018cvk,Cappiello:2019qsw,Ema:2018bih}, we assume the galactic HECRs are uniformly and isotropically distributed in a thick disk with a radius of $R=10$ kpc and a half-height $h=1$ kpc. 
We use the observed local CR spectrum shown in Fig.~29.8 of Ref.~\cite{Tanabashi:2018oca}, which follows a broken power over a wide energy range up to $10^{21}$ eV\footnote{Note that there are different measurements in Fig. 29.8 of Ref.~\cite{Tanabashi:2018oca}, which could differ from each other by a factor of 2. We use the Grigorov and CASA-MIA data below PeV and the mean values of all the measurements above.}.
For simplicity, we only consider contributions from protons and scale the CR flux above $\sim 100$~TeV in \cite{Tanabashi:2018oca} by a factor of 0.3 to match the well-measured local CR proton flux below $\sim 100$~TeV \cite{Boschini:2017fxq}.
We neglect the small velocity of DM and simply assume them at rest initially.

The differential DM flux per solid angle $\Omega$ (in unit of ${\rm GeV^{-1}~cm^{-2}~s^{-1}~sr^{-1}}$) up-scattered by CRs and arriving at Earth is given by a line-of-sight (l.o.s.) integral \cite{Bringmann:2018cvk,Cappiello:2019qsw,Ema:2018bih}
\begin{align}
\Phi_\chi^{\rm MW} (E_\chi,\Omega) \equiv \frac{d^2N_\chi}{dE_\chi d\Omega} (E_\chi,\Omega) = \int_{\rm l.o.s.} d\ell \int dE_p \frac{\rho_\chi(r)}{m_\chi}\Phi_p(E_p) D_{p\chi}(E_p, E_\chi) \sigma_{p\chi}, \label{eq:flx_chi_gal} \end{align}
where $\Phi_p \equiv d^2N_p/(dE_p d\Omega)$ is the differential flux of CRs (in unit of ${\rm GeV^{-1}~cm^{-2}~s^{-1}~sr^{-1}}$) and
$\rho_\chi$ is the halo DM mass density. For $\rho_\chi$, we take the Navarro-Frenk-White (NFW) profile \cite{Navarro:1995iw,Navarro:1996gj} with a scale radius $r_s=20$~kpc, 
normalized to the local DM density $\rho_0=0.3\gev~{\rm cm}^{-3}$ where the Sun is located at $r=8.2~{\rm kpc}$ from the galactic center (GC). 
We note that the single-component DM scenario is assumed in this work so that the DM local density is entirely made of $\chi$.
The transfer function $D_{p\chi}(E_p, E_\chi)$ encodes 
the produced spectrum of up-scattered DM with total energy $E_\chi$ from scattering with a CR proton of energy $E_p$.
Assuming that the scattering is elastic and isotropic in the center-of-mass (c.m.) frame, 
\begin{align}
 D_{p\chi}(E_p, E_\chi) = \frac{\Theta[E^{\rm max}_\chi(E_p)-E_\chi]}{E_\chi^{\rm max}(E_p)},
\label{eq:D}
\end{align}
with 
\begin{align}
 E_\chi^{\rm max} \simeq \frac{E_p }{1+(m_p^2+m_\chi^2)/(2m_\chi E_p)},
 \label{eq:emax}
\end{align}
for $E_p\gg m_p,m_\chi$ in the energy range of concern in this work.

Note that the scattering c.m. energy $\sqrt{s}$ could be as large as $10$~GeV for $m_\chi \gtrsim 100$~keV. 
In this case, inelastic scattering can occur to allow multiple-meson production. 
We argue, however, that the transfer function in Eq.~(\ref{eq:D}) should still provide a good estimation of the up-scattered $E_\chi$. The reason is that the inelastic scatterings are dominated by resonant channels at low energies and DIS at high energies, and both can be effectively described by a 2 $\to$ 2 scattering process (see, e.g., Refs. \cite{Mucke:1999yb,Formaggio:2013kya} for nucleon excited by photon or neutrino).
Without otherwise stated, we stick to the expression in Eq.~(\ref{eq:D}) throughout this work. 

\begin{figure*}[htbp]
\begin{centering}
\subfloat[]{
\includegraphics[width=0.51\textwidth]{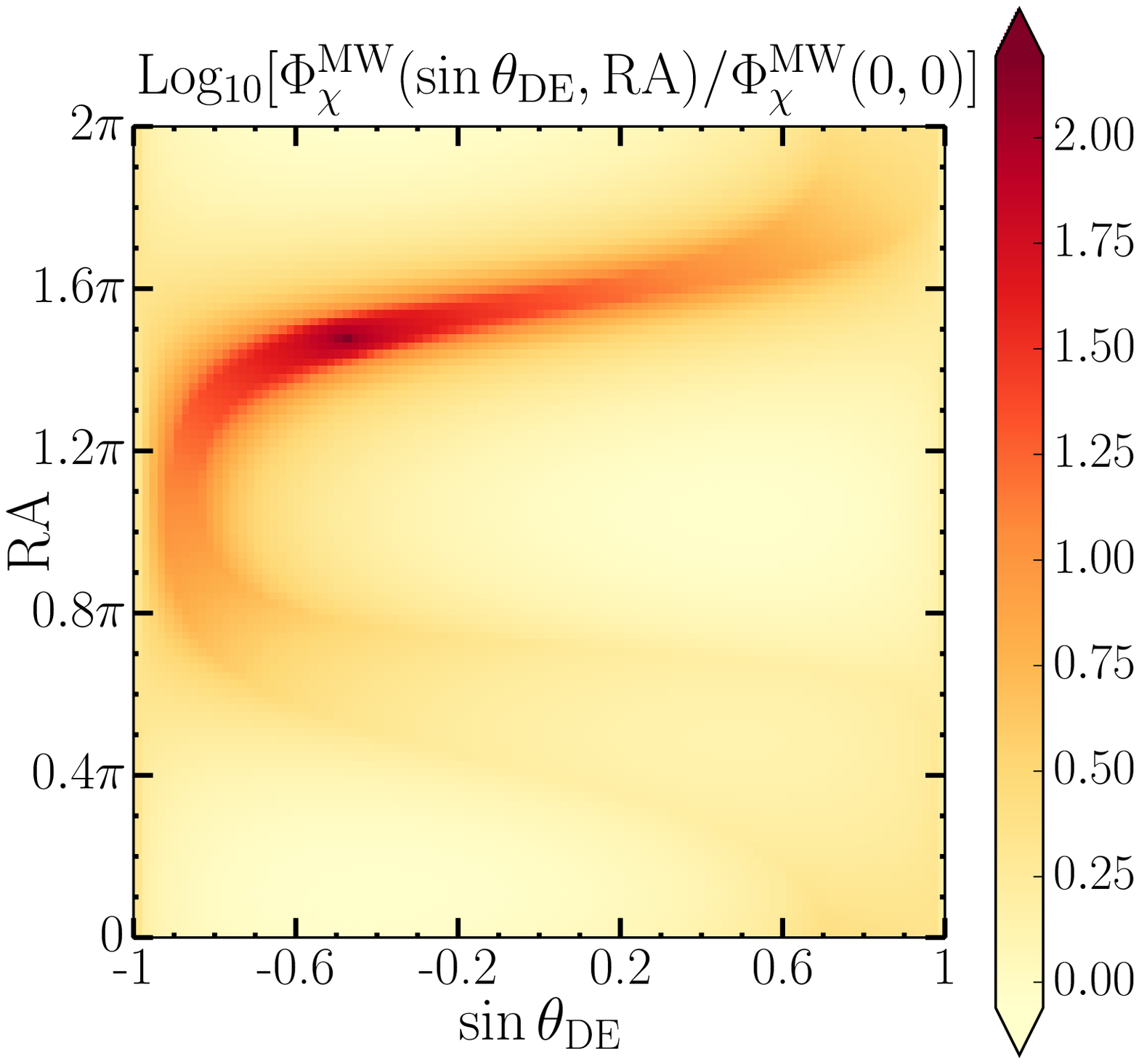}
}
\subfloat[]{
\includegraphics[width=0.51\textwidth]{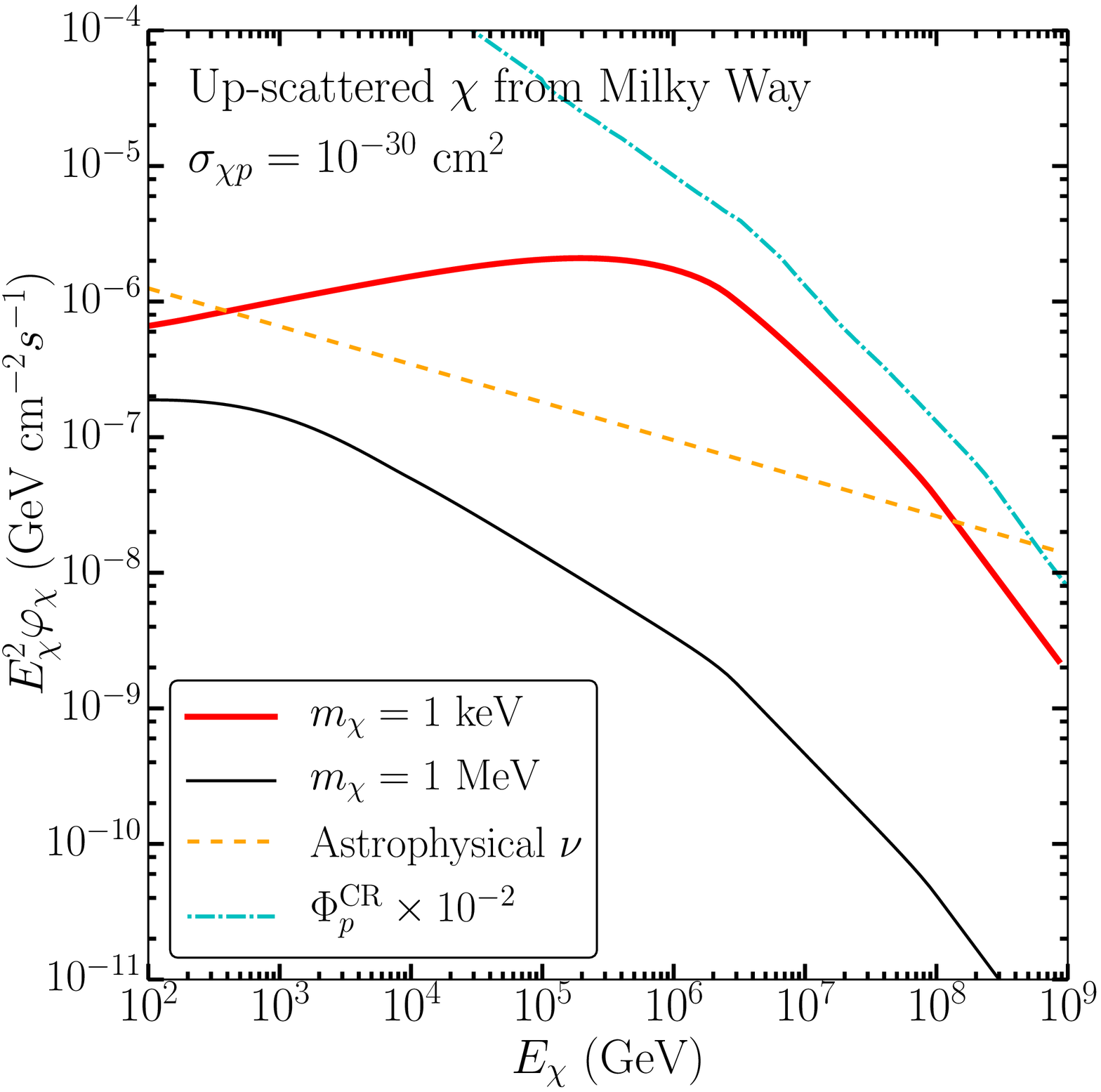}
}
\caption{(a) The angular distribution of DM flux 
$\Phi_\chi^{\rm MW}(E_\chi,\Omega)$ 
up-scattered by CR protons in the Milky Way 
as a function of the 
spherical equatorial coordinates $\sin\theta_\texttt{DE}$ and $\texttt{RA}$, normalized by the value at $\sin\theta_\texttt{DE}=0$ and $\texttt{RA}=0$.
(b) The up-scattered DM energy spectrum $E_\chi^2\varphi_\chi(E_\chi)\equiv E_\chi^2 \int d\Omega \Phi_\chi^{\rm MW}(E_\chi,\Omega)$ assuming a DM--proton scattering cross-section $\sigsip=10^{-30}$~cm$^2$. 
The thick-red-solid and thin-black-solid lines are for $m_\chi=1\kev$ and $m_\chi=1\mev$, respectively.
Also shown are the CR proton flux (cyan dash-dotted line) and the best-fit HE astrophysical neutrino flux measured by IceCube (orange dashed line).
}
\label{fig:DM_flux}
\end{centering}
\end{figure*}

The l.o.s. integral in Eq.~(\ref{eq:flx_chi_gal}) can be easily computed in the galactic coordinate system. To compare with the IceCube data, we re-express the differential DM flux in the spherical equatorial coordinate, and use right ascension (\texttt{RA}) and declination ($\theta_\texttt{DE}$) to describe the arriving angles of DM.
In Fig.~\ref{fig:DM_flux}, we show the resulting angular distribution of up-scattered DM flux in panel (a), as well as their energy spectra (integrated over $\Omega$) in panel (b) for cases with $\sigsip=10^{-30}$~cm$^2$ and $m_\chi=1$~keV (red line) and 1~MeV (black line). 
The angular distribution shown in panel (a) is normalized by 
the flux at $\texttt{RA}=\sin\theta_\texttt{DE}=0$ and shows clearly that most of the up-scattered DM come from directions within the reddish band which spans over the disk wherein CRs are confined. 
In particular, the angular distribution peaks at
the direction of the GC with $\texttt{RA}=1.48\pi$ and $\sin\theta_\texttt{DE}=-0.48$. 
Note that this distribution is independent of the chosen $m_\chi$ and $\sigsip$ because they can be factored out.

The energy spectra, however, show strong dependence on $m_\chi$.
Using Eqs.~\eqref{eq:flx_chi_gal}--\eqref{eq:emax}, it is straightforward to show that the DM energy spectrum $\int d\Omega \Phi_\chi^{\rm MW} \propto E_\chi^{-(\gamma_p+1)/2}$ for $E_\chi\ll (m_p^2+m_\chi^2)/(2m_\chi)$, 
where $\gamma_p$ is the CR spectral index characterizing the CR spectra $\Phi_p\propto E_p^{-\gamma_p}$ around $E_p\sim E_\chi$.
On the other hand, for $E_\chi\gg (m_p^2+m_\chi^2)/(2m_\chi)$, 
$\int d\Omega \Phi_\chi^{\rm MW} \propto E_\chi^{-\gamma_p}$.
The break at $E_\chi\sim (m_p^2+m_\chi^2)/(2m_\chi)$ originates from the maximum DM energy $E_\chi^{\rm max}$ that can be accelerated by CR with energy $E_p$ shown in Eq.~\eqref{eq:emax}.
This behavior is best illustrated by the red curve in Fig.~\ref{fig:DM_flux}(b) which breaks at $E_\chi\sim 0.5$~PeV for $m_\chi=1$~keV (see the rescaled $\Phi_p$ shown in the same panel for comparison).
Comparing the two curves with $m_\chi=1$~keV and 1~MeV, Fig.~\ref{fig:DM_flux}(b) also shows that for a larger $m_\chi$, the resulting DM flux is smaller as the halo DM number density is inversely proportional to $m_\chi$.
On the other hand, the difference between them becomes smaller at $E_\chi\lesssim 0.5$~GeV due to the break of spectrum for $m_\chi=1$~keV.

\subsection{Extragalactic HE DM flux}\label{sec:EGDM}

The up-scattered DM of extragalactic origin can come from different sources. 
First, the accelerated DM produced within all other galaxies by the CRs trapped therein by the same mechanism discussed in Sec.~\ref{sec:GDM} can arrive the Earth as a diffuse source. 
Second, the extragalactic HECRs propagating in the intergalactic space can also up-scatter the cosmic DM that have not collapsed to form galaxies.
Both components will arrive as isotropic and diffuse HE DM sources and are indistinguishable.
We argue, however, that the first extragalactic component is always subdominant compared to the galactic flux. For example, if we assume all galaxies produce HE DM particles at a same rate, $\dot{\mathcal{N}}_\chi$, as the Milky Way, and neglect all the effects due to cosmic evolution and expansion. The energy-integrated flux arriving at the Earth from other galaxies is  
\begin{align}
\int dE_\chi \Phi_\chi & \sim \int^{r_{\rm max}}_0 \frac{\dot{\mathcal{N}}_\chi n_{\rm galaxy}}{4\pi r^2} 4\pi r^2 dr \nonumber \\ 
& \sim 0.1 \frac{\dot{\mathcal{N}}_\chi}{(10~{\rm kpc})^2} \left[\frac{n_{\rm galaxy}}{0.1~({\rm Mpc})^{-3}}\right] 
\left[\frac{r_{\rm max}}{10~{\rm Gpc}}\right].
\label{eq:EGDM1est}
\end{align}
Taking $n_{\rm galaxy}\sim 0.1~{\rm Mpc}^{-3}$ which is the typical number density for small galaxies between $10^7$ and $10^{10}$~$M_\odot$~\cite{Conselice:2016xx}, 
Eq.~\eqref{eq:EGDM1est} indicates that the HE DM flux 
from this component is likely at least a factor of 10 smaller 
than the galactic flux $\sim \dot{\mathcal{N}}_\chi/({\rm a~few~kpc})^2$.
Although the higher star forming activities for galaxies at high red-shift may enhance this flux by a factor of a few (see below), it can partly be compensated by the cosmic expansion as well as the expected fewer amount of DM in smaller galaxies. As all these are subject to the large uncertainties, we only focus on the HE DM produced in the intergalactic space hereafter, which, we show later, can have a higher flux compared to the galactic component.

Similar to the galactic case, the flux of up-scattered DM (in unit of ${\rm GeV^{-1}~cm^{-2}~s^{-1}~sr^{-1}}$) from collision between the diffuse intergalactic HECRs and the comic DM background can be expressed as       
\begin{align}\label{eq:EGDMflux}
\Phi_\chi^{\rm EG}(E_\chi)  = \int dE_p \int_0^\infty dz  \sigma_{p\chi} n_\chi^0 (1+z)^4 F(E_p,z)D_{p\chi}(E_p,E_\chi(1+z))\left|\frac{cdt}{dz}\right|, 
\end{align}
where $|dt/dz|^{-1}=H_0(1+z)[\Omega_\Lambda + \Omega_M(1+z)^3]^{1/2}$.
For the cosmological parameters, we take $H_0=70~{\rm km~s^{-1}~Mpc^{-1}}$, $\Omega_\Lambda=0.7$, and $\Omega_M=0.3$.
The current cosmic DM number density is estimated\footnote{Note that the actual amount of cosmic DM in the intergalactic space may vary by $\sim$ a factor of two as a fraction of them are locked into the galaxies, see e.g., Ref.~\cite{Masaki:2012xx}.} to be 
$n_\chi^0 \simeq 10^{-6}~ (\gev/m_\chi)~{\rm cm}^{-3}$ 
and the corresponding DM density at redshift $z$ is enhanced by a factor of $(1+z)^3$.
Another enhancement factor of $(1+z)$ in Eq.~\eqref{eq:EGDMflux} accounts for the compression of the energy scale due to the cosmic expansion.

For the diffuse CR flux $F(E_p, z)$ at redshift $z$, since the attenuation due to the interaction with the cosmic microwave background (CMB) or the diffuse IR/UV background is negligible for protons with energy below $\sim 10^{18}$ eV~\cite{Berezinsky:2002nc}, 
it is simply given by
\begin{align}
 F(E_p, z) = \frac{c}{4\pi} \int_{z}^\infty S(z') \frac{1+z'}{1+z} F_{\rm inj}\Big(\frac{1+z'}{1+z}E_p\Big)  \left|\frac{dt}{dz^\prime}\right| dz',
 \label{eq:Fpz}
\end{align}
where $F_{\rm inj}(E_p)$ is the injected CR spectrum per unit time and per unit co-moving volume (in unit of ${\rm GeV^{-1}~cm^{-3}~s^{-1}}$), and $S(z)$ accounts for the cosmological evolution of the sources.
As the spectrum of the extragalactic HECRs at our interested energy range $E_p\lesssim 10^{18}$~eV is unknown, below we consider sources of ultra-high-energy CRs (UHECRs), namely, CRs with energies above $\sim 10^{18}$ eV, and extrapolate the injected spectrum down to $E_p \sim$ PeV following a broken power law. It should also be pointed out that in Eq.~(\ref{eq:Fpz}) we have neglected the time for CRs to escape from the galaxies that produce them.
For CR protons at $\sim 10$ GeV, the escape time is about 10--100 Myr \cite{Yanasak_2001}. Assuming that the escape time decreases as $E_p^{-\delta}$ with $\delta \sim 0.3$--0.6 \cite{Strong:2007nh}, PeV CR protons can diffuse out the galaxies within a few Myr, which is much shorter than the cosmological time scales. 

The origin of the UHECRs still remains a mystery. Particularly, depending on the mass composition, there are different scenarios that are able to explain the observed features of UHECRs, including an ankle at energy around $5 \times 10^{18}$~eV as well a sharp drop of the spectrum at $3 \times 10^{19}$~eV \cite{Aloisio:2012ba,Aloisio:2017qoo}. 
Assuming a pure proton composition for the UHECRs, the ankle and the cutoff can be explained naturally 
by the pair production and photohadronic processes of protons on the CMB (the so-called \textit{dip model}) \cite{Berezinsky:2002nc}. 
An alternative model, \textit{the mixed composition model} \cite{Aloisio:2013hya}, interprets the ankle as a transition between two different types of components, a steep spectrum at low energy with a proton-dominated composition and a hard one at high energy with increasing fractions of heavy isotopes. 
The flux suppression at the highest energies is due to the photo-disintegration of heavy nuclei on the CMB and/or the maximum accelerated energy \cite{Aloisio:2013hya}. 
Despite this ambiguity in modeling the UHECRs, both models tend to indicate a very similar injected proton-dominated flux around $10^{18}$ eV relevent to this work. Thus, we only consider the dip model from Refs.~\cite{Aloisio:2015ega,AlvesBatista:2019rhs} in the rest of this section. 

\begin{figure}[htbp]
\begin{centering}
\includegraphics[width=0.65\textwidth]{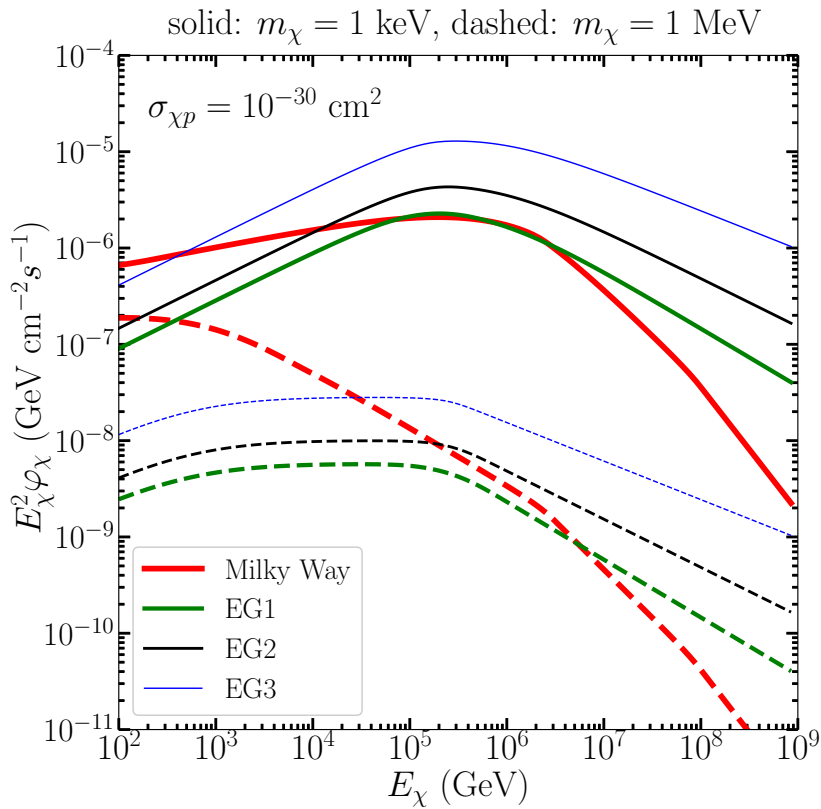}
\caption{The DM energy spectrum up-scattered by extragalactic CR protons. 
The three cases EG1 (green lines), EG2 (black lines), and EG3 (blue lines) assuming different cosmological evolution history of the CR source injection (see text for details) are shown together with 
the DM spectrum up-scattered by CRs within the Milky Way (red lines). 
The solid and dashed lines are for $m_\chi=1$~keV and  $m_\chi=1\mev$, respectively. The DM--proton scattering cross-section is fixed to $10^{-30}$~cm$^2$.
}
\label{fig:EG_spectrum}
\end{centering}
\end{figure}

We assume that the injected protons follow a broken power law, a hard spectrum at low energies followed by a soft one at high energies:  
\begin{align}
 F_{\rm inj}(E_p) =
 N_0 \left\{ \begin{array}{c} E_c^{-\gamma+2}E_p^{-2},\;\;\; E_p < E_c, \\ E_p^{-\gamma},\quad\quad\qquad  E_p \ge E_c. \end{array} \right. 
 \label{eq:finj}
\end{align}
The normalization constant $N_0$ and the spectral index $\gamma$ are tuned to fit the observed UHECRs taking into account the CR propagation in the intergalactic medium.
The breaking energy $E_c$ is a free parameter mainly constrained by CR observation.
However, a $E_c$ that is as low as a few GeV would require an energy budget beyond known astrophysical sources~\cite{Berezinsky:2002nc}
Throughout this work, we simply take $E_c = 10^{15}$ eV.
For the dip model, we consider the following three cases with different spectral indexes and cosmological evolution history, which all fit the UHECR data well \cite{AlvesBatista:2019rhs}. 
\begin{itemize}
 \item Model EG1: $\gamma=2.6$ and there is no source evolution, i.e., $S(z)=1$.
 \item Model EG2: $\gamma=2.5$ and $S(z)$ follows the star formation rate (SFR),
  \begin{align} 
S(z) = \left\{ \begin{array}{cc} (1+z)^{3.4}, & z\le 1, \\ 2^{3.7}\times(1+z)^{-0.3}, & 1<z\le 4, \\ 2^{3.7}\times 5^{3.2} \times (1+z)^{-3.5}, & z>4. \end{array} \right.
\end{align}
 \item Model EG3: $\gamma=2.4$ and a stronger cosmological evolution of the sources,
 \begin{align}
S(z) = \left\{\begin{array}{cc} (1+z)^5, & z\le 1.7, \\ 2.7^{5}, & 1.7<z\le 2.7, \\ 2.7^5 \times 10^{2.7-z}, & z>2.7. \end{array} \right.
\end{align}
 \end{itemize}

The corresponding source emissivities at redshift $z=0$,
\begin{align}
{\cal L}_0  = \int_{10^{16}~{\rm eV}}^{10^{23}~{\rm eV}} dE_p F_{\rm inj}(E_p),         
\end{align}
required to reproduce the observed data are ${\cal L}_0 \simeq 1.5 \times 10^{46}$, $6\times 10^{45}$, and $3 \times 10^{45}~{\rm erg~Mpc^{-3}~yr^{-1}}$ for the three different models, respectively. From the values of ${\cal L}_0$, the normalization constant $N_0$ for $F_{\rm inj}$ can be determined [see Eq.~(\ref{eq:finj})].

Fig.~\ref{fig:EG_spectrum} shows the HE DM energy spectra predicted by the three models for $\sigsip=10^{-30}$~cm$^2$ and $m_\chi=1$~keV and $1$~MeV.
Similar to the galactic case, the extragalactic DM flux also shows a spectral break at $E_\chi\sim (m_p^2+m_\chi^2)/(2m_\chi)$.
For all three models, the extragalctic HE DM flux can be comparable or a factor of a few higher than the galactic component. 
Model EG3 predicts a largest flux, followed by the model EG2 and EG1, due to the larger contribution at redshift $z\simeq 1$--2.

\section{Analysis and results}\label{sec-result}

The HE DM particles up-scattered by collision with galactic and extragalactic CRs discussed in the previous section can generate cascade events in IceCube, which are indistinguishable from the ones induced by neutrinos. 
We use the IceCube 2-year medium-energy starting events (MESEs) \cite{Aartsen:2014muf}, i.e., those with deposited energies above 1 TeV and interaction vertex contained within the detector, to constrain the up-scattered DM signals.
There are in total 383 events, among which, 278 are cascades and 105 are tracks. 
If produced by neutrinos, the cascades are due to the neutral current (NC) interaction of all flavors as well as the charged current (CC) interactions of $\nu_{e,\tau}$, $\bar\nu_{e,\tau}$. On the other hand, the tracks are associated with $\mu^\pm$ produced from $\nu_{\mu,\tau}$ and $\bar\nu_{\mu,\tau}$ CC interactions. In principle, the 6-year high-energy starting events (HESEs) \cite{Aartsen:2013jdh,Aartsen:2014gkd,Kopper:2015vzf,Kopper:2017zzm} could also be used for our study. In this work we use the MESE sample because more detailed effective areas associated with the MESEs are publicly available.

Since the HE DM scattering with targets can mimic the cascade events, one can obtain a conservative limit on $\sigma_{\chi p}$ by requiring the total cascade events caused by DM not exceeding the observed IceCube data, as done in Refs.~\cite{Bringmann:2018cvk,Ema:2018bih,Cappiello:2019qsw}. 
In this work, we take a more comprehensive approach
by explicitly considering the cascades and tracks from atmospheric muons/neutrinos as well as from astrophysical neutrinos.
This leaves less room for the DM signals and thus would result in tighter constraints on $\sigma_{\chi p}$. As the galactic HE DM flux contains a strong angular dependence (see Fig.~\ref{fig:DM_flux}), we also take into account the arrival-angular information of the IceCube events.

In what follows, we will take a binned likelihood analysis to probe/constrain the DM signals. 
We first discuss how to estimate the event numbers from the atmospheric background, the astrophysical neutrinos, as well as the up-scattered DM. 
We also discuss how to choose the bins for the deposited energy $E_{\rm d}$ and the arriving angles at IceCube. 
We then construct the binned likelihood function and obtain the limits on $\sigma_{\chi p}$ using the up-scattered DM from within the Milky Way and from the intergalactic space.        

\subsection{Backgrounds: atmospheric background and astrophysical neutrinos}\label{sec:background}

Backgrounds to astrophysical neutrino searches are dominated by atmospheric muons and neutrinos produced from CR showers.
A purely atmospheric explanation for the IceCube data has been already excluded at $8\sigma$ \cite{Ahlers:2018fkn}, indicating the existence of astrophysical neutrinos. 
However, to probe the up-scattered DM signals with IceCube, the atmospheric muons/neutrinos and astrophysical neutrinos all need to be considered as background sources. 

We directly use the estimated background of atmospheric muons/neutrinos from the IceCube paper analysing a 2-year-data set of MESEs \cite{Aartsen:2014muf}, where the distributions in declination angles and deposited energies are available. 
As for astrophysical neutrinos, we rely on the flux measured by the throughgoing muons (TGM) \cite{Aartsen:2016xlq,Stettner:2019tok}. 
Assuming a single power-law for neutrino energy $E_\nu$ ranging from a few TeV to 10 PeV, the best-fit flux per flavor for the TGM events is \cite{Stettner:2019tok}
\begin{align}
\Phi_{\nu+\bar\nu}(E_\nu) = N_{\nu} \left(\frac{E_\nu}{100~{\rm TeV}}\right)^{-\gamma_{\nu}} \cdot 10^{-18}~{\rm GeV^{-1}~cm^{-2}~s^{-1}~sr^{-1}},  \label{eq:fmu}
\end{align}  
where 
\begin{equation}
N_{\nu} = 1.44^{+0.25}_{-0.24} ~~{\rm and}~~\gamma_{\nu}=2.28^{+0.08}_{-0.09}. \label{eq:astro_pars}
\end{equation}
Note that the above analysis assumes that fluxes arriving at IceCube are equally partitioned among all neutrino flavors, which naturally arises from the averaged oscillation of neutrinos produced by pion decay from a distant astrophysical source. Detailed studies also showed that the current IceCube data are consistent with the flavor equipartition scenario from charged $\pi$ decay \cite{Aartsen:2015ivb,Palomares-Ruiz:2015mka}, though a recent study from \cite{Palladino:2019pid} favours neutron decay as the production mechanism. Unless otherwise stated, we always take the flavor equipartition assumption and use the diffuse per-flavor neutrino flux in Eq.~(\ref{eq:fmu}) to estimate their contribution to the 2-year MESE data set. If the HE neutrinos detected at IceCube are truly from neutron decay, more cascades will be expected from the astrophysical neutrinos, which only makes our derived constraints in Sec.~\ref{sec:constraint} even stronger.  

\subsection{Events in each bin from the background and the up-scattered DM}

We use the tables of effective areas from Ref.~\cite{Aartsen:2014muf} for the MESE sample to calculate the event numbers from the astrophysical neutrinos.
The effective areas encode the probability of detecting neutrinos with an incoming energy, $E_\nu$, an arriving declination angle, $\theta_\texttt{DE}$, and a deposited energy, $E_{\rm d}$, at IceCube.
They also depend on the neutrino flavor, $f$, the interaction channel, $c$, and the event topology, $t$. In the effective area tables provided by IceCube, $E_\nu$ ranges from $10^2$ GeV to $10^9$ GeV and is divided into 100 bins with equal size in logarithmic scale. 
Similarly, 20 equal-size bins between $E_{\rm d}=10^2$ GeV and $10^8$ GeV are taken logarithmically.  For $\theta_\texttt{DE}$, $\sin\theta_\texttt{DE}$ is divided into 10 equal bins with a bin size of 0.2. Using the effective areas, the expected event numbers of astrophysical neutrinos in given bins of $E_\nu$, deposited energy, $E_{\rm d}$, and the declination, $\sin\theta_\texttt{DE}$ can be obtained (see details in Appendix~\ref{sec:events_nu}). 

To perform a binned likelihood analysis of the observed IceCube events, one can simply take the same bins of $E_{\rm d}$ and $\sin\theta_\texttt{DE}$ as those in the IceCube effective area tables, which would, however, lead to a very limited statistics.
We have scrutinized several ways of energy binning and found that the derived bounds are not sensitive to the choice of binning in $E_d$. Therefore, we simply use one $E_{\rm d}$-bin for our analysis. 
For the angular binning, 
we take the same bins for $\sin\theta_\texttt{DE}$ as in the effective area tables. 
As the $\texttt{RA}$ values of IceCube MESEs are not publicly available, we only use one single bin for $\texttt{RA}$.
Due to the strong angular dependence of the galactic HE DM flux, a further binning in $\texttt{RA}$ in addition to $\sin\theta_\texttt{DE}$ could in principle help enhance the sensitivity.   
In our analysis, we consider all the 31 IceCube MESEs with $25~{\rm TeV}\le E_{\rm d}\le 100$ PeV, $-1 \le \sin\theta_\texttt{DE}\le 0$, and $0 \le \texttt{RA}\le 2\pi$.
Note that we only include the downgoing events in order to maximize the ratio of DM signals to the astrophysical neutrino signals, since the Earth attenuation is typically stronger for DM with the $\sigsip$ range of concern.
Combining the events computed with the original bins listed in the effective area tables, we can obtain the expected event number from the astrophysical neutrinos in our chosen bins. 

To compute the number of HE DM induced cascade events at IceCube, we can first obtain a similar effective area table for DM detection at IceCube as for the HE neutrinos. The attenuation of downgoing DM due to $\chi$--$p$ scattering when traversing the ice shell needs be taken into account.
For this purpose, we estimate the effective area for HE DM by scaling the effective area for cascade production from the NC interaction of $\nu_e$ using the same bins. Similarly, the event numbers of DM in our chosen bins can be obtained. More relevant details are presented in Appendix~\ref{sec:events_dm}. 

\begin{figure}[htbp]
\begin{centering}
\includegraphics[width=0.65\textwidth]{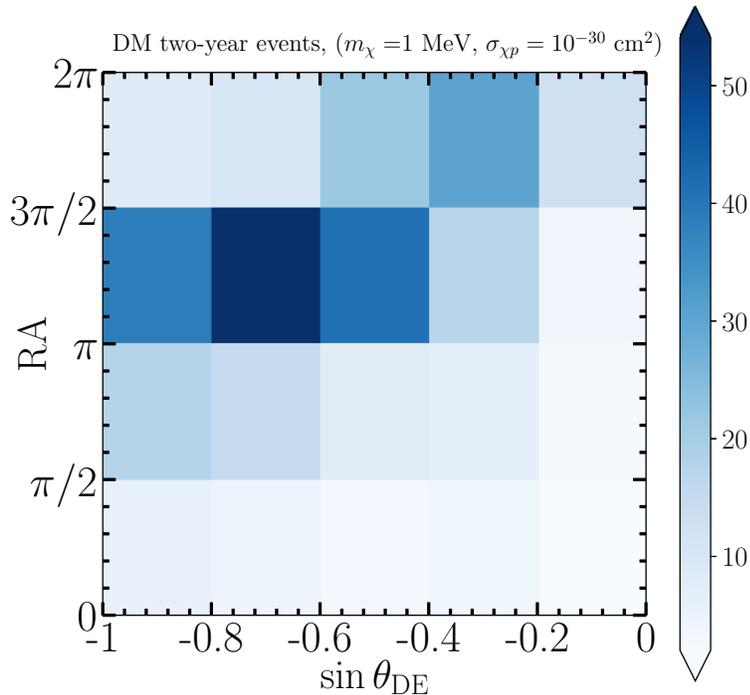}
\caption{The two dimensional histogram showing the number of HE DM events that would be detected by IceCube in each bin on the ($\sin\theta_{\texttt{DE}}$,~$\texttt{RA}$) plane with $25~{\rm TeV}\le E_{\rm d}\le 100$ PeV, assuming  
$m_\chi=1\mev$ and $\sigsip=10^{-30}$~cm$^2$. 
Only the up-scattered DM by the galactic CRs are considered for this plot. The bin containing the most events includes the direction towards the galactic center.
}
\label{fig:NDM}
\end{centering}
\end{figure}

Taking a total exposure time of $T=641$~days and summing over the $E_{\rm d}$ bins, 
we show in Fig.~\ref{fig:NDM} the angular distribution of the galactic HE DM induced down-going cascade events ($\sin\theta_{\texttt{DE}}<0$) for a benchmark case with $m_\chi=1$~MeV and $\sigsip=10^{-30}$~cm$^2$.
Although we use one \texttt{RA}-bin for our likelihood analysis in the next section, here we choose to show the expected numbers of DM events in five $\sin\theta_{\texttt{DE}}$-bins and four \texttt{RA}-bins. As the effective areas are independent of $\texttt{RA}$, the expected numbers in each bin are calculated using Eq.~\eqref{eq:dmevent} by replacing the factor $2\pi$ with $\pi/2$.
In contrast to the fairly isotropic distribution of the HE neutrino induced cascades, it shows strong angular dependence on both the RA and DE due to the reason discussed in Sec.~\ref{sec:GDM}.
Comparing with Fig.~\ref{fig:DM_flux}, 
the angular bin  having the most events ($-0.8 \leq \sin\theta_\texttt{DE}<-0.6$ and $\pi\leq \texttt{RA}< 3\pi/2$) does not contain the direction pointing to the GC. This is due to the effect of attenuation when DM traversing through the ice which suppresses the event number more for less negative $\sin\theta_\texttt{DE}$.

\subsection{Constraints on the DM--nucleon scattering cross-section at TeV--PeV}\label{sec:constraint}

In this section, we perform a binned likelihood analysis by combining the atmospheric muon/neutrino background and the astrophysical HE neutrinos, together with the predicted DM events produced by both the galactic and extragalactic sources, to derive constraints on $\sigsip$.
Note that as the c.m. energy of $\chi$--$p$ collision required to accelerate DM to $\sim$ PeV is lower than the corresponding c.m. energy for detecting a DM particle with energy $\sim$ PeV at IceCube by a factor of $m_p/m_\chi$, the involved $\sigsip$ can generally be different at these two sites due to its energy dependence.
Below, we assume that $\sigsip$ is energy-independent following Refs.~\cite{Bringmann:2018cvk,Ema:2018bih,Cappiello:2019qsw} for most cases.
We further discuss how the excluded region 
varies when an energy-dependent cross-section is considered for both the galactic and extragalactic HE DM.

We define the likelihood from our binned data as
\begin{equation}
\mathcal{L}(m_\chi, \sigma_{\chi p})= \max_{N_\nu,\gamma_\nu} \left[ 
e^{-\frac{1}{2}(\chi_{N_\nu}^2+\chi_{\gamma_\nu}^2)} \times
\prod_{k,t} \frac{e^{-(s_{kt}+b_{kt})} (s_{kt}+b_{kt})^{N_{kt}}}{N_{kt}!}
\right],
\label{eq:likelihood}    
\end{equation}
where $s_{kt}$ and $b_{st}$ are the expected event numbers from HE DM and the background (including the atmospheric muons/neutrinos and the astrophysical neutrinos), respectively, in the $k$-th bin of $\sin\theta_\texttt{DE}$\footnote{Note that for the likelihood analysis we always take one single bin for both $E_{\rm d}$ and $\texttt{RA}$.}, and $N_{kt}$ is the corresponding observed event number at IceCube in the same bins. 
The parameter $t$ stands for the event type, which includes track and cascade.
As discussed in Sec.~\ref{sec:background}, we rely on the flux measured using the TGMs to estimate the event numbers from the astrophysical neutrinos. To account for the flux uncertainty, we introduce the nuisance parameters $N_\nu$ and $\gamma_\nu$, corresponding to the normalization and the spectral parameter of the astrophysical neutrino flux, respectively [see Eq.~(\ref{eq:astro_pars})].
We take Gaussian distributions for the nuisance parameters, i.e., $\chi^2_{N_\nu} \equiv (N_\nu - \bar N_\nu)^2/\sigma_{N_\nu}^2$ and $\chi^2_{\gamma_\nu} \equiv (\gamma_\nu -\bar \gamma_\nu)^2/\sigma_{\gamma_\nu}^2$, with $\bar N_\nu = 1.44$, $\bar\gamma_\nu=2.28$, $\sigma_{N_\nu} = 0.25$, and $\sigma_{\gamma_\nu}=0.09$ [see Eq.~(\ref{eq:astro_pars})]. We vary them to maximize the likelihood, as shown in Eq.~\eqref{eq:likelihood}.
Note that both $N_\nu$ and $\gamma_\nu$ have asymmetric uncertainties, and we take the larger ones for $\sigma_{N_\nu,\gamma_\nu}$ for conservative studies.

\begin{figure}[htbp]
\begin{centering}
\subfloat[]{
\includegraphics[width=0.51\textwidth]{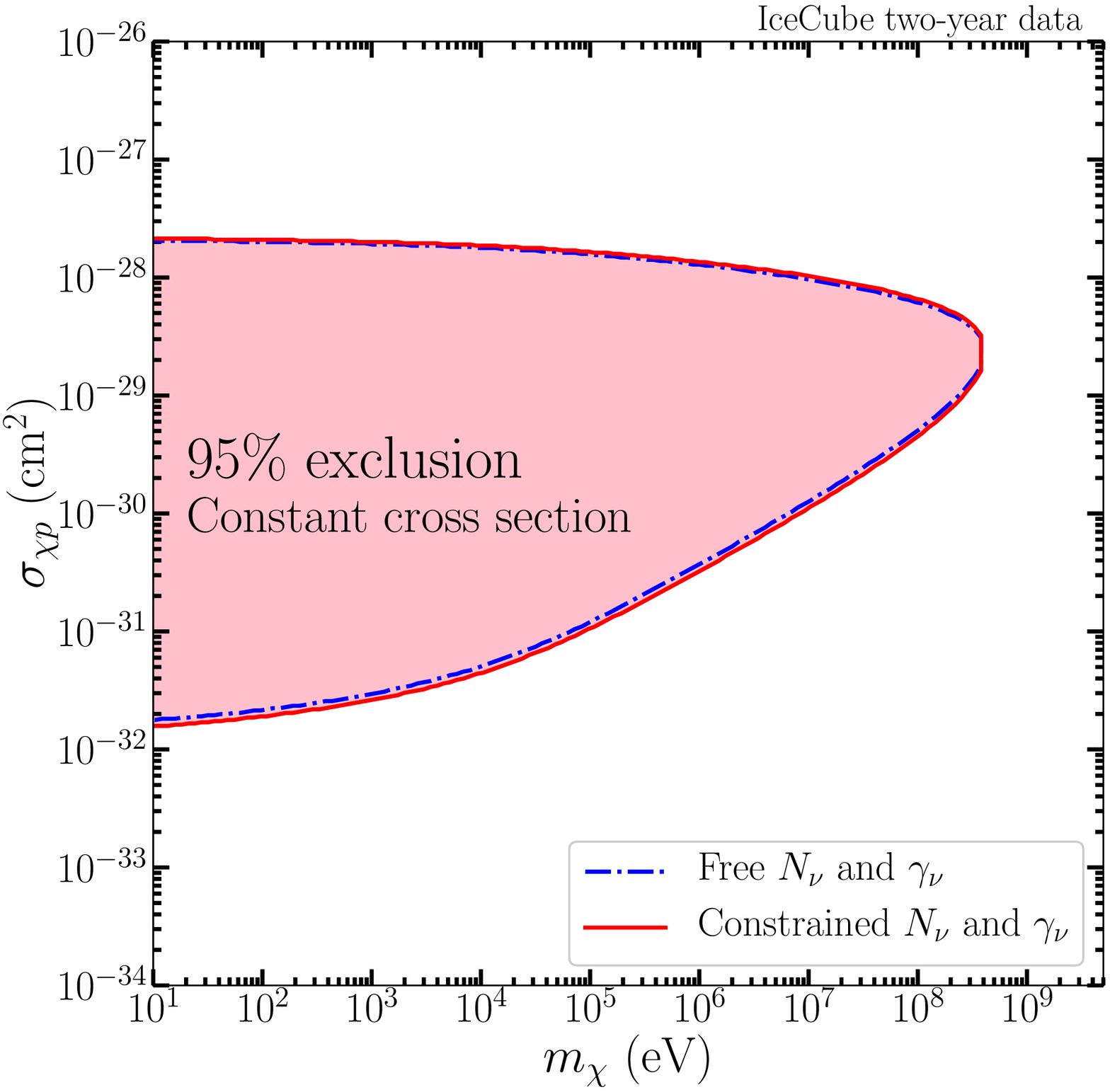}
}
\subfloat[]{
\includegraphics[width=0.51\textwidth]{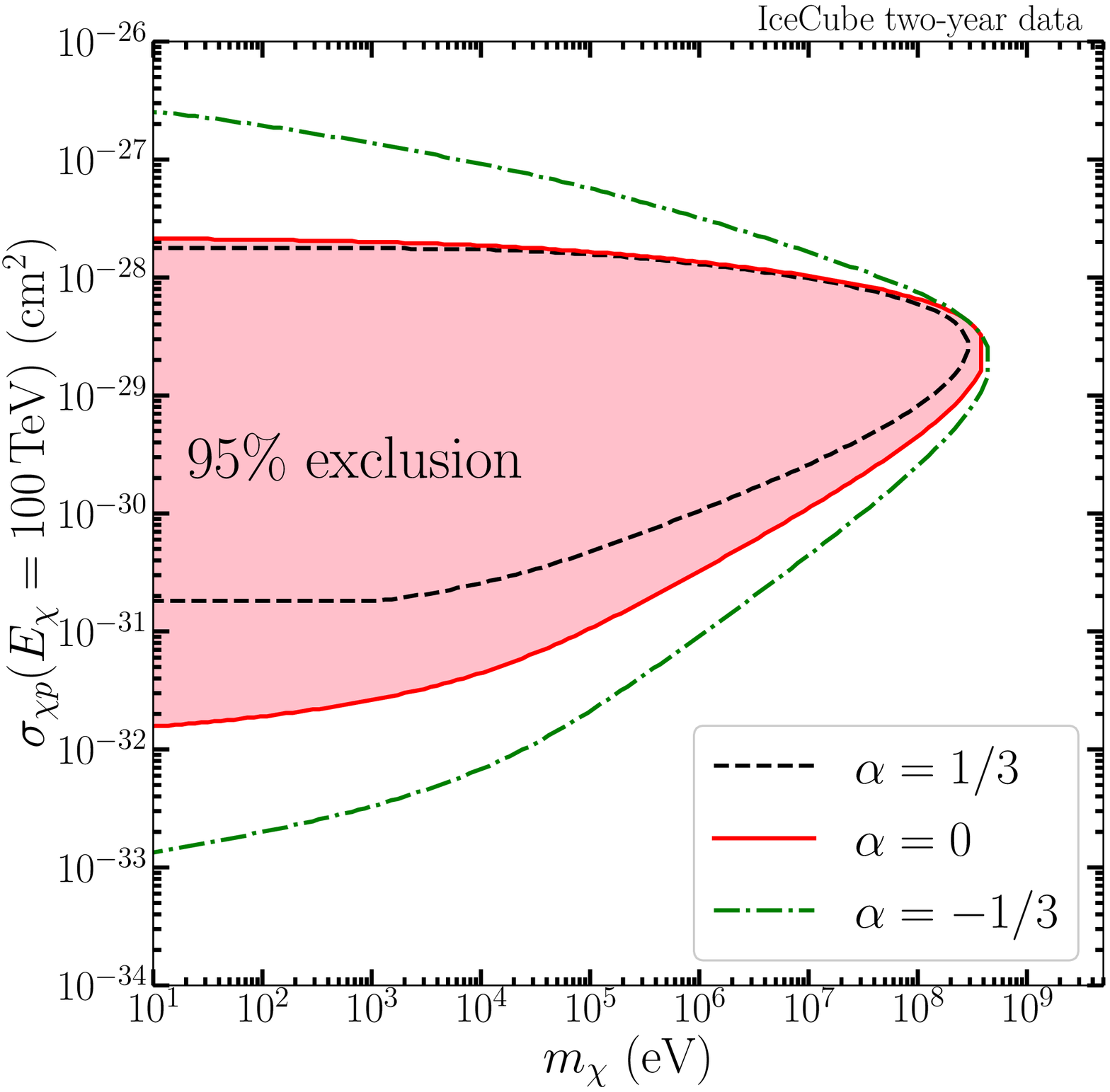}
}
\caption{The $95\%$ exclusion limit projected on the ($m_\chi$, $\sigsip$) plane considering up-scattered DM by galactic CRs.      
In panel (a), the excluded regions are derived assuming a constant $\sigsip$ with different assumptions on astrophysical neutrino flux parameters $N_{\nu}$ and $\gamma_{\nu}$
[see Eq.~\eqref{eq:astro_pars}].
The red solid line is derived assuming $N_\nu$ and $\gamma_\nu$ following Gaussian distributions given by the best fit to the TGM events at the IceCube (see Sec.~\ref{sec:background}).
The blue dash-dotted line is obtained without assuming any knowledge for $N_\nu$ and $\gamma_\nu$.
In panel (b), we show the derived exclusion limits for $\sigsip$ at $E_\chi=100\tev$ 
when an energy-dependent $\sigsip\propto E_\chi^\alpha$ is taken, see Eq.~\eqref{eq:egydep}. 
All three lines here are derived taking $N_{\nu}$ and $\gamma_{\nu}$ distribution given by TGM events.
}
\label{fig:exc}
\end{centering}
\end{figure}

We define the one-sided 95\% confidence level (CL) (equivalent to a two-sided 90\% CL) for the excluded region by finding where the $-2\Delta\ln (\mathcal{L})=-2\ln[\mathcal{L}(m_\chi, \sigma_{\chi p})/\mathcal{L}(m_\chi, 0)]=2.71$.
Fig.~\ref{fig:exc} shows the $95\%$ excluded regions on the ($m_\chi$, $\sigsip$) plane for
cases considering the galactic HE DM component.
Similar to those derived in Refs.~\cite{Bringmann:2018cvk,Ema:2018bih,Cappiello:2019qsw},
for a given $m_\chi$, all the excluded regions have not only
lower bounds below which the $\sigsip$ is too small to produce enough HE DM flux and to interact with target at the IceCube, but also upper bounds above which the attenuation effect takes place to prevent enough HE DM reaching the detector.
The excluded regions shrink with increasing $m_\chi$ and vanish at $m_\chi \gtrsim$~GeV as the DM number density inside the galactic halo decreases with $m_\chi$.

We show in Fig.~\ref{fig:exc}(a) the excluded regions (the red solid curve) using the likelihood defined in Eq.~\eqref{eq:likelihood}.
In addition, we test the robustness of the excluded region by relaxing the TGM constraint on the astrophysical neutrinos flux and show the resulting blue dash-dotted curve in the same panel. Specifically, we remove the penalty term, $\exp[-\frac{1}{2}(\chi_{N_\nu}^2+\chi_{\gamma_\nu}^2)]$, from Eq.~\eqref{eq:likelihood} and project the excluded region onto the ($m_\chi$, $\sigsip$) plane by freely varying $N_\nu$ and $\gamma_\nu$.
Comparing the blue dash-dotted curve with the red solid curve, it shows that the derived lower-bound only weakly depends on the assumption of astrophysical neutrino flux.
We also note that the upper bound at $\sigsip\sim 2\times 10^{-28}$~cm$^{-2}$ for $m_\chi\lesssim 1$~MeV is almost independent of this assumption. 
This is simply because all the downgoing DM are absorbed by the ice shell before reaching the IceCube detector located at 1,450~m below the surface when $\sigsip$ is larger than this value.
Based on these, we can confidently exclude a wide range of $2\times 10^{-32}~{\rm cm}^2\lesssim \sigsip \lesssim 
2\times 10^{-28}~{\rm cm}^2$ for $m_\chi\lesssim 10$~keV, as well as an extended region for $m_\chi$ up to $\sim 1$~GeV when considering the galactic HE DM component only and assuming a constant $\sigsip$ at all relevant energies.

In Fig.~\ref{fig:exc}(b), we show how an energy-dependent $\sigsip$ affects the derived bounds.
Taking a simple power-law dependence of $\sigma_{\chi p}$ on $E_\chi$,
\begin{align}
\sigsip(E_\chi)=\sigma_{\chi p,0}\times \left(\frac{E_\chi}{ 100~{\rm TeV}}\right)^\alpha, 
\label{eq:egydep}
\end{align}
with $\sigma_{\chi p ,0}$ corresponding to $\sigsip$ at $E_\chi=100$~TeV in the rest-frame of the proton,
Fig.~\ref{fig:exc}(b) shows that with a positive (negative) value of $\alpha=1/3$ ($\alpha=-1/3$), the excluded region shrinks (enlarges) substantially.
This is because for light DM considered here, $E_\chi$ in the proton-rest-frame at the acceleration site is smaller than that at the IceCube by a factor of $\sim m_p/m_\chi$ as mentioned in the beginning of this subsection. 
Thus, when we fix $\sigsip$ for different $\alpha$ at $E_\chi=100$~TeV, a positive (negative) $\alpha$ results in
smaller (larger) values of lower bounds on $\sigsip$
because the DM flux $\Phi_\chi^{\rm MW}$ scales proportionally to $\sigsip$ at the acceleration site [see Eq.~\eqref{eq:flx_chi_gal}].
The same reason also explains why the excluded range in $m_\chi\sim$~GeV around $\sigsip\simeq 10^{-29}$~cm$^2$ slightly shrinks (extends) to a smaller (larger) value when $\alpha$ is positive (negative).
For the upper bounds, taking $\alpha=1/3$ results in similar values to the case with $\alpha=0$ due to the similarly strong attenuation at $E_\chi\gtrsim$~PeV.
However, for $\alpha=-1/3$, the upper bounds extend to substantially larger value of $\sigsip$ because for $E_\chi\gtrsim$~PeV, it requires larger values of $\sigsip$ at 100~TeV for most DM to be absorbed before reaching the IceCube.

\begin{figure}[htbp]
\begin{centering}
\subfloat[]{
\includegraphics[width=0.51\textwidth]{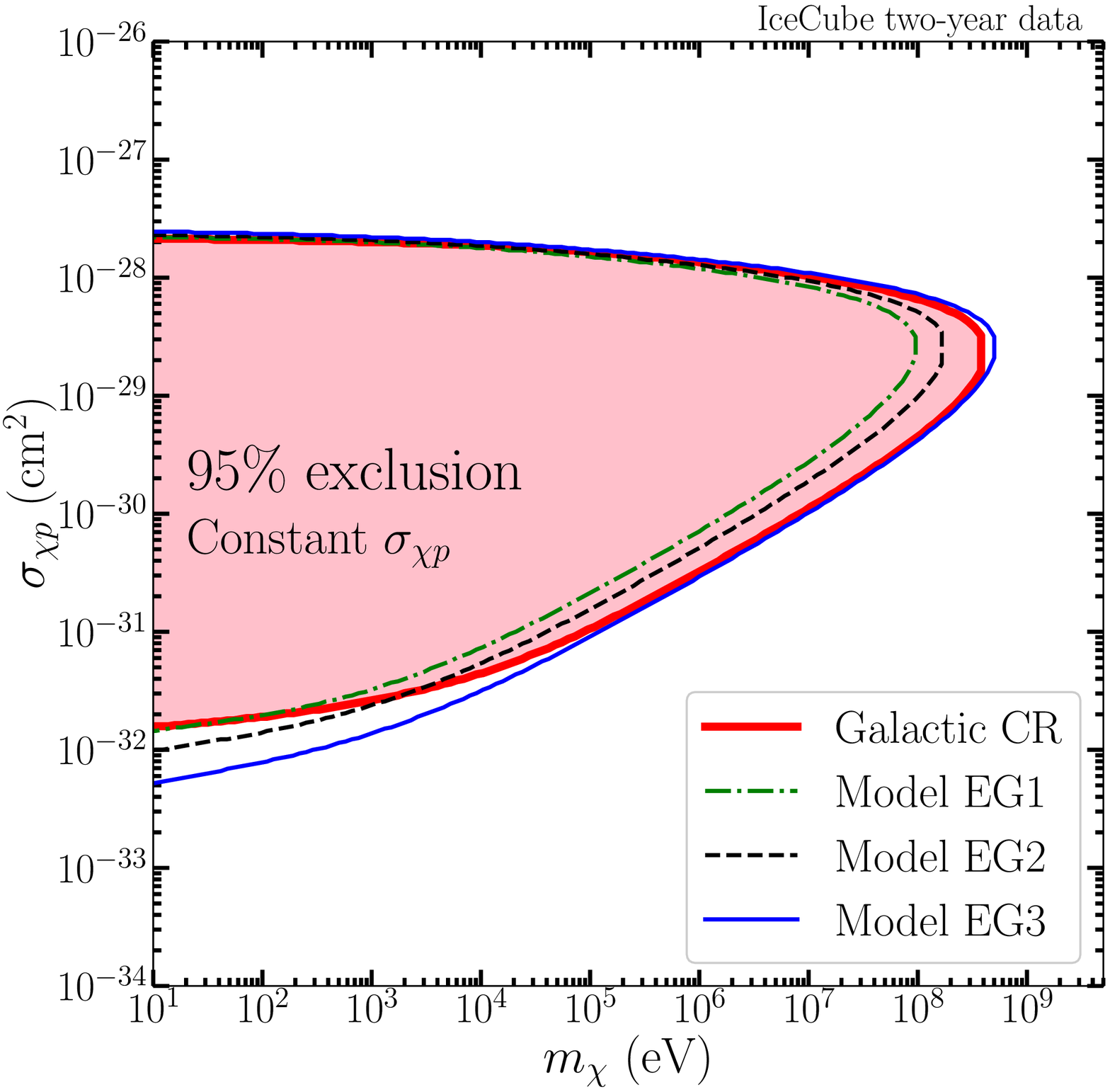}
}
\subfloat[]{
\includegraphics[width=0.51\textwidth]{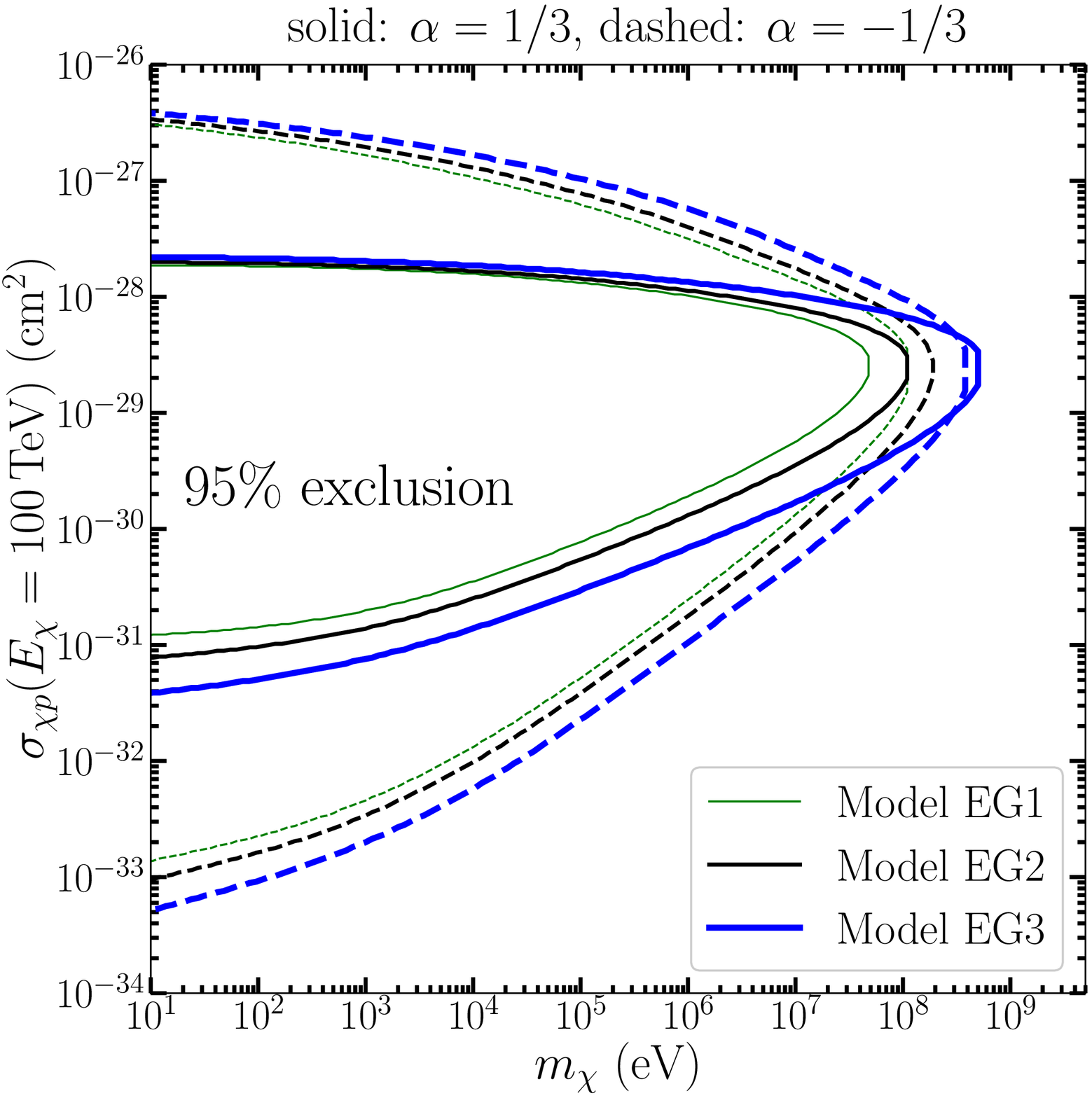}
}
\caption{The $95\%$ exclusion limits projected on the ($m_\chi$, $\sigsip$) plane derived when considering up-scattered DM by extragalactic CRs. 
In panel (a), the exclusion contours using different extragalactic CR models EG1, EG2, and EG3 are shown by the
green dash-dotted, black dashed, and blue solid line, respectively.
Also shown is the red contour obtained by considering the galactic component (see Fig.~\ref{fig:exc}).
In panel (b), we show the corresponding exclusion limits assuming an energy-dependent $\sigsip$ as in Fig.~\ref{fig:exc}(b) for $\alpha=1/3$ and $-1/3$. 
}
\label{fig:EGCR}
\end{centering}
\end{figure}

We now turn our attention to the HE DM component accelerated by the extragalactic HECRs (see Sec.~\ref{sec:EGDM}).
Fig.~\ref{fig:EGCR} shows the excluded regions derived with three extragalactic models
considered in Sec.~\ref{sec:EGDM} for cases with a constant $\sigsip$ in panel (a) and with energy-dependent $\sigsip$ in panel (b).
All three models yield bounds comparable to that derived from the galactic DM component. 
Note that although the angle-integrated extragalactic DM fluxes from all three models in the relevant energy range exceed that of the galactic component (see Fig.~\ref{fig:EG_spectrum}), 
only the model EG3 results in a slightly larger excluded region. 
This is simply due to the extragalactic HE DM flux being isotropic, unlike the galactic component which has a strong angular dependence.
Taking an energy-dependent $\sigsip$ results in similar changes to the excluded regions here as for the case of the galactic DM discussed above.

\subsection{Comparison with existing bounds}\label{sec:comparison}

\begin{figure}[htbp]
\begin{centering}
\includegraphics[width=0.65\textwidth]{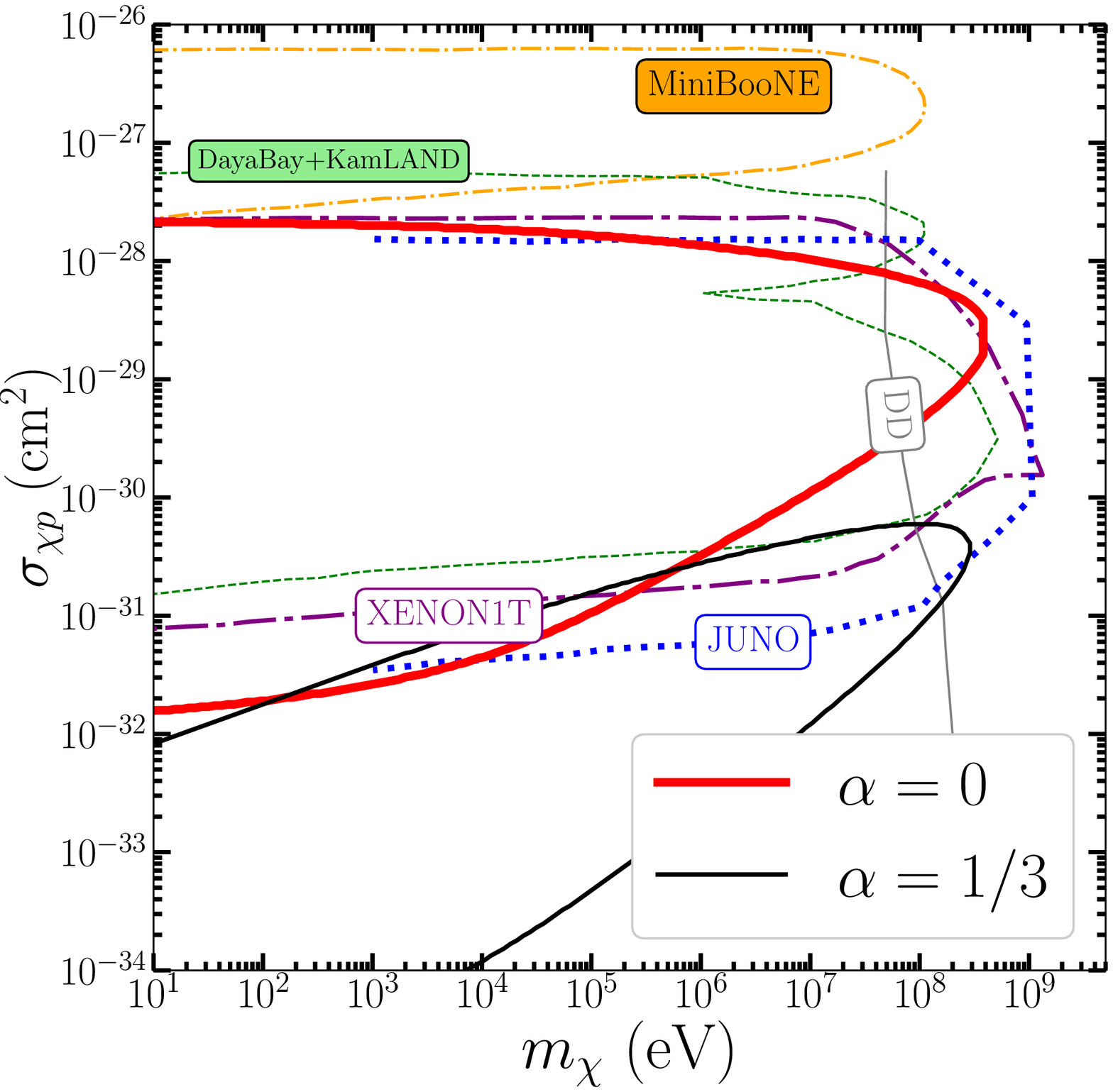}
\caption{Comparison of the excluded regions derived in this work with others. 
Our limits for energy independent ($\alpha=0$) and energy dependent ($\alpha=1/3$) cases 
are shown by red-thick-solid line and black-thin-solid line, respectively. 
The cross-section of the energy dependent ($\alpha=1/3$) case has been extrapolated to nonrelativistic regime for DM velocity of  $10^{-3}c$, comparable to the local DM velocities.
The orange dash-dotted (labeled by MiniBooNE) and purple long dashed lines (labeled by XENON1T) are given by Ref.~\cite{Bringmann:2018cvk}. 
The green thin dashed (labeled by DayaBay+KamLAND) and the blue dotted lines (labeled by JUNO) are the current and projected limits from Ref.~\cite{Cappiello:2019qsw}.
The grey curve labeled by DD is the limit given by the DM direct detection experiments CRESST~\cite{Abdelhameed:2019hmk}.
}
\label{fig:comparison}
\end{centering}
\end{figure}

The constraints on $\sigsip$ obtained in Sec.~\ref{sec:constraint} are very unique because they probe the HE DM interaction with nucleons at $E_\chi\sim$~TeV--PeV.
To compare our bounds with those derived from other considerations such as the DM direct detection for nonrelativistic DM or the CR up-scattered DM with $E_\chi\lesssim$~GeV, proper modeling of how $\sigsip$ changes with $E_\chi$ is needed.
This can depend on the exact nature of interaction of DM with nucleon.
Here, we simply take the $\sigsip(E_\chi)$ given in Eq.~\eqref{eq:egydep} and extrapolate the excluded region shown in Fig.~\ref{fig:exc} to the nonrelativistic regime where the DM velocity $v_{\chi}= 10^{-3}~c$, 
for the case with $\alpha=1/3$.

Fig.~\ref{fig:comparison} shows the results (red and black solid curves) together with different bounds shown in Fig.~7 of Ref.~\cite{Cappiello:2019qsw}. 

The bounds labeled by MiniBooNE and XENON1T are given by Ref.~\cite{Bringmann:2018cvk}. The ones labeled by DayaBay+KamLAND and JUNO are the current and projected limits from Ref.~\cite{Cappiello:2019qsw}. The grey curve labeled by DD is the limit given by the DM direct detection experiments CRESST~\cite{Abdelhameed:2019hmk}.
For the constant $\sigsip$ case, the bound that we derived here is comparable to those obtained by considering CR up-scattered DM in energy range of $\sim$ MeV--GeV~\cite{Bringmann:2018cvk,Cappiello:2019qsw}
for $m_\chi\gtrsim 1$~MeV but weaker than the projected sensitivity with JUNO.
However, for $m_\chi\lesssim 1$~MeV, we improve the existing bounds by roughly one order of magnitude.
When we consider energy dependent $\sigsip$ with $\alpha=1/3$,
the excluded region gets shifted to smaller $\sigsip$ by orders of magnitudes (cf.~Fig.~\ref{fig:exc}), allowing to exclude $\sigsip$ down to $\sigsip\ll 10^{-32}$~cm$^2$ for $m_\chi\lesssim 1$~MeV.
We note that if the energy dependence of $\sigsip$ resembles the behavior of neutrino--nucleon interaction, we can expect that the excluded region extrapolated to low energy can be pushed to even smaller $\sigsip$.

\section{Summary and discussion}\label{sec-summary}

In this paper, we have investigated the scenario that light DM particles ($m_\chi \lesssim 1$~GeV) can be up-scattered to $\sim 10$~TeV--PeV energy range by the galactic and extragalactic HECRs and its consequences at the HE neutrino detectors like the IceCube.
For a DM--nucleon cross-section $\sigma_{\chi p}\sim 10^{-30}~{\rm cm}^{-2}$ at $E_\chi\sim 100$~TeV (in the rest frame of the nucleon), 
the resulting HE DM fluxes can be comparable to or larger than that of HE neutrinos detected at the IceCube.
As these HE DM only produce cascade events in IceCube, differently from HE neutrinos that can also produce tracks, we derive for the first time strong constraints on DM--nucleon cross-section at the energy range of $E_\chi\sim$~PeV for a wide range of $\sigsip$ and $m_\chi$ using the IceCube 2-year MESE sample.

As the energy of DM in the rest frame of nucleons at the acceleration site can differ from that at the detection site by a factor $m_p/m_\chi$, we also explored how an energy-dependent $\sigsip$ affects the derived bounds taking a simple power-law of $\sigsip(E_\chi)$ beyond the constant $\sigsip$ assumption taken in previous work investigating the up-scattered DM by galactic CRs of $\lesssim$~GeV~\cite{Bringmann:2018cvk,Ema:2018bih,Cappiello:2019qsw}.
Taking a positive (negative) power-law index of $1/3$ in the energy dependence of $\sigsip$, we showed that the derived bounds shrink (expand) up to a factor of $\sim 10$, depending on $m_\chi$.
If naively extrapolating the energy dependence down to low $E_\chi$ of the order of MW halo DM, the corresponding constraints on $\sigsip$ can be pushed to even lower values if $\sigsip$ decreases as $E_\chi$.
Thus, future work combining existing low-energy limits and our new constraints at high energy with a consistent modeling of the energy dependence of $\sigsip$ can be valuable and will be further pursued.

In this study, we take uniformly distributed galactic CRs and the NFW DM profile. Considering the full spatial dependence of the galactic CR distribution or a different DM halo profile may affect the predicted signal number by a factor of a few. As the signal number depends on $\sigma_{\chi p}^2$, the uncertainties in limits from modeling of the CRs and the DM profile are suppressed by a square root.      

We focus on the energy region of TeV--PeV relevant for IceCube. Since the fluxes of the CRs and HE DM decrease rapidly with a spectral index of $\sim -3$, we expect less stringent limits when considering experiments like ANITA \cite{Gorham:2008dv,Allison:2018cxu}, which is sensitive to higher energy region around $10^{18}$ eV.
We also note that the up-scatter HE DM signals may possibly alleviate the spectral discrepancy between the IceCube track and cascade datasets \cite{Kopper:2017zzm,Aartsen:2016xlq,Denton:2018aml,Stettner:2019tok}, although our analysis using the current data only provides a slight improvement. We expect that as increasing amount of data being collected by the IceCube, its future upgrade to IceCube-Gen2, and the upcoming full implementation of KM3NeT, not only further improved constraints can be obtained, hints for light DM signals may also be revealed. 
Our work here thus highlights that the current and future HE neutrino detectors can probe independently the nature of DM--nucleon interaction at energy scales inaccessible to other experiments.    

Throughout this work, we have not considered the inelasticity of DM--nucleon collision, which can produce indirect signals like HE gamma-rays or HE neutrinos from meson-decay.
As the flux of such signals only depends on the $\sigsip$ at the collision site of the CRs and DM, the results will not depend on the particular assumption of the energy dependence of $\sigsip$ across a larger energy range and can therefore provide further independent information.
Moreover, unlike the constraints derived in this paper which has no sensitivity for large $\sigsip$ due to the attenuation effect, we expect that $\sigma_{\chi p}$ above certain values will be all excluded by considering these secondary signals.
However, production of mesons can depend on the specific DM models.
Constraints based on these will be published in a separate work.

\begin{acknowledgments}
We thank Martin Spinrath and Wei-Chih Huang for providing useful comments to this manuscript.
We also thank an anonymous referee for useful suggestions that helps improve the presentation of this work.
G.~G. and M.-R.~W. acknowledge support from the Academia Sinica by Grant No.~AS-CDA-109-M11. Y.-L.~S.~Tsai
was funded in part by the Chinese Academy of Sciences Taiwan Young Talent Programme
under Grant No.~2018TW2JA0005.
M.-R.~W. acknowledges support from the Ministry of Science and Technology, Taiwan under Grant No.~108-2112-M-001-010, and the Physics Division, National Center of Theoretical Science of Taiwan.
\end{acknowledgments}

\appendix
\section{Event numbers from the astrophysical neutrinos and the HE DM}

\subsection{Events from the astrophysical neutrinos}
\label{sec:events_nu}

 The effective area tables for HE neutrinos, $A_{{\rm eff},f,c,t}^{ijk}$, are provided by IceCube \cite{Aartsen:2014muf} to calculate the event numbers in bins of the deposited energy, $E_{\rm d}$, and arriving angle, $\theta_\texttt{DE}$, for a given flux in bins of the incoming neutrino energy, $E_\nu$, with $i,j,k$ being the indices of $E_\nu$-bin, $E_{\rm d}$-bin, and $\sin\theta_\texttt{DE}$-bin, respectively. We would like to note that the effective areas correspond to the medium energy starting events (MESEs) with deposited energies down to 1 TeV, and for consistency purposes, we use the 2-year MESE data sample for our analysis.
The effective areas also depend on the neutrino flavor, $f$, interaction channel, $c$, including the neutral current (NC) scattering on nucleon and charged current (CC) scattering on nucleon, as well as the Glashow resonance (GR) for $\bar\nu_e$ scattering with electrons, and the event topology, $t$, which could be cascade or track.
 Note that we have simply assumed a perfect angular resolution for the IceCube detector, i.e., the reconstructed angle is the same as the incoming angle.
 This assumption only affects our results minorly because the angular bin taken in our analysis has a comparable size as the angular uncertainty, which, for cascades, is about 10$^\circ$--15$^\circ$. 
 
As in the effective area tables, we divide $\sin\theta_\texttt{DE}$ into 10 equal bins with a bin size of 0.2. The energy $E_\nu$ in [$10^2$ GeV, $10^9$ GeV] and $E_{\rm d}$ in [$10^2$ GeV, $10^8$ GeV] are divided into 100 and 20 bins with equal bin size in logarithmic scale. 
The expected event number of cascades or tracks in each given bin of deposited energy, $E_{\rm d}$, and $\sin\texttt{DE}$ is given by       
\begin{align} 
 n_{\nu,t}^{jk} = \sum_{i,f,c} [\Phi_{\nu+\bar\nu}(E^i_\nu)\Delta E_\nu^i] \times A_{{\rm eff},f,c,t}^{ijk} \times  \Delta\sin\theta_\texttt{DE} \times 2\pi \times T,     
 \label{eq:nnu}
\end{align}
with $\Delta\sin\theta_\texttt{DE}=0.2$ and $T=641$ days during which the IceCube 2-year data were taken. We always assume that astrophysical neutrinos are equally partitioned among all neutrino/antineutrino flavors, and $\Phi_{\nu+\bar\nu}$ is the neutrino flux per flavor including neutrinos and antineutrinos. Note that the effective areas provided by IceCube are different for neutrinos and antineutrinos for given flavor, and we use the averaged values for $A_{{\rm eff},f,c,t}^{ijk}$ in Eq.~(\ref{eq:nnu}).
   
As discussed in the main text, we bin $E_{\rm d}$ and $\sin\theta_\texttt{DE}$ differently from that in the effective area tables to obtain better constraints.
The corresponding event numbers in the new bins can be simply obtained by merging the events given in Eq.~(\ref{eq:nnu}).

\subsection{Events from the up-scattered DM}
\label{sec:events_dm}

To compute the number of HE DM induced cascade events at the IceCube, their attenuation due to $\chi$--$p$ scattering when traversing through the Earth/ice shell needs to be taken into account in the same way as the astrophysical neutrinos.
For this purpose, we estimate the effective area for HE DM at the IceCube, $A_{{\rm eff},\chi}^{ijkm}$, by scaling the effective area for cascade production from the NC interaction of $\nu_e$ in Eq.~\eqref{eq:nnu}:

\begin{equation}
A_{{\rm eff},\chi}^{ijk}=A_{{\rm eff},\nu_e,{\rm NC,cascade}}^{ijk}
  \frac{\langle e^{-\tau_{\chi}} \sigma_{\chi p}\rangle^{ik}}
  {\langle e^{-\tau_{\nu_e}} \sigma_{\nu_e p}^{\rm NC}\rangle^{ik}},
\end{equation}
where $i,j,k$ are the indices of bins as in Eq.~(\ref{eq:nnu}) and
$\langle e^{-\tau_{\alpha}} \sigma_{\alpha p}\rangle^{ik}$ is the averaged value of the product of the survival probability for a given angular and energy bin and the cross-section,   
\begin{equation}
\langle e^{-\tau_{\alpha}} \sigma_{\alpha p} \rangle^{ik} =  \frac {\int_{\sin\theta_\texttt{DE}^{k,{\rm min}}}^{\sin\theta_\texttt{DE}^{k, {\rm max}}} d\sin\theta_\texttt{DE} \int_{0}^{2\pi} d\texttt{RA} ~e^{-\tau_\alpha(E^i, \sin\theta_\texttt{DE})} \sigma_{\alpha p}(E^i) \Phi_\alpha(E^i, \Omega)}
{\int_{\sin\theta_\texttt{DE}^{k,{\rm min}}}^{\sin\theta_\texttt{DE}^{k, {\rm max}}} d\sin\theta_\texttt{DE} \int_{0}^{2\pi} d\texttt{RA} ~\Phi_\alpha(E^i,\Omega)
},  
\label{eq:atten_ave}
\end{equation}
with the optical depth given by 
\begin{equation}
 \tau_{\alpha}(E, \sin\theta_\texttt{DE})= N_A \cdot X(\sin\theta_\texttt{DE}) \cdot  \sigma_{\alpha p}^{\rm tot}(E),
\end{equation}
for $\alpha=\nu_e$ or $\chi$.
In the above equation, $N_A$ is the Avogadro constant,
$X(\sin\theta_\texttt{DE}) = \int \rho d\ell$ is the column depth that DM or neutrino traverses inside the Earth/ice shell
from the declination angle $\sin\theta_\texttt{DE}$.
As we consider a 3 km-thick ice layer with a constant density $\sim 1~{\rm g~cm^{-3}}$, we only need to consider attenuation in the ice shell for downgoing DM events studied in our analysis (almost no attenuation for neutrinos in the considered energy range). For computing $X(\sin\texttt{DE})$, we simply neglect the finite size of the IceCube detector and consider a single depth of 1,450 m for the whole detector. In principle, fully taking into account the finite size effect of the detector can slightly affects our results. The correction due to this should be smaller than astrophysical uncertainties, e.g., the spatial dependence of the CRs and the DM halo profile.
Note that the atmosphere density is too low and the related attenuation effect is neglected.
$\sigma^{\rm tot}_{\nu_e p} = \sigma^{\rm CC}_{\nu_e p} + \sigma^{\rm NC}_{\nu_e p}$ and $\sigma^{\rm tot}_{\chi p}=\sigma_{\chi p}$ are the total cross-section of $\nu_e$--proton scattering taken from Ref.~\cite{Connolly:2011vc} and the DM--proton cross-section, respectively. 
Since both the extragalactic HE DM flux and the astrophysical neutrino flux are assumed to be isotropic, the integration over $\texttt{RA}$ in Eq.~\eqref{eq:atten_ave} is trivial and can be cancelled out.
Given the above effective area of the HE DM, we can then compute the corresponding DM induced cascade number at the IceCube 
in each given bin with the deposited energy $E_{\rm d}^j$, $\sin\theta^k_\texttt{DE}$:
\begin{align}
 n_{\chi}^{jk} = \sum_i \left (\Phi_\chi^{ik}\Delta E_\chi^i\right) \times A_{{\rm eff},\chi}^{ijk} \times  \Delta\sin\theta_\texttt{DE} \times 2\pi \times T.     
 \label{eq:dmevent}
\end{align}
Similarly to the HE neutrino, the event number of HE DM in the bins chosen for our binned analysis
can be obtained by combining the events in Eq.~(\ref{eq:dmevent}).   

\bibliographystyle{JHEP}

\providecommand{\href}[2]{#2}\begingroup\raggedright\endgroup

\end{document}